\documentclass[revtex4]{emulateapj}

\begin{document}

\title{Spectroscopic investigation of a reionized galaxy overdensity at z=7} 

\author{M. Castellano\altaffilmark{1}, L. Pentericci\altaffilmark{1}, E. Vanzella\altaffilmark{2}, F. Marchi\altaffilmark{1}, A. Fontana\altaffilmark{1}, P. Dayal\altaffilmark{3}, A. Ferrara\altaffilmark{4}, A. Hutter\altaffilmark{5}, S. Carniani\altaffilmark{6,7},  S. Cristiani\altaffilmark{8,9}, M. Dickinson\altaffilmark{10}, S. Gallerani\altaffilmark{4}, E. Giallongo\altaffilmark{1}, M. Giavalisco\altaffilmark{11}, A. Grazian\altaffilmark{1}, R. Maiolino\altaffilmark{6,7}, E. Merlin\altaffilmark{1}, D. Paris\altaffilmark{1}, S. Pilo\altaffilmark{1}, P. Santini\altaffilmark{1}}

\altaffiltext{1}{INAF - Osservatorio Astronomico di Roma, Via Frascati 33, I - 00040 Monte Porzio Catone (RM), Italy}
\altaffiltext{2}{INAF - Osservatorio Astronomico di Bologna, Via Ranzani 1, I - 40127, Bologna, Italy}
\altaffiltext{3}{Kapteyn Astronomical Institute, University of Groningen, P.O. Box 800, 9700, AV Groningen, The Netherlands}
\altaffiltext{4}{Scuola Normale Superiore, Piazza dei Cavalieri 7, I-56126 Pisa, Italy}
\altaffiltext{5}{Swinburne University of Technology, Hawthorn, VIC 3122, Australia}
\altaffiltext{6}{Cavendish Laboratory, University of Cambridge, 19 J. J. Thomson Ave, Cambridge CB3 0HE, UK }
\altaffiltext{7}{Kavli Institute for Cosmology, University of Cambridge, Madingley Road, Cambridge CB3 0HA, UK}
\altaffiltext{8}{INAF - Osservatorio Astronomico di Trieste, Via G. B. Tiepolo 11, I-34143 Trieste, Italy}
\altaffiltext{9}{INFN - National Institute for Nuclear Physics, via Valerio 2, I-34127,  Trieste, Italy}
\altaffiltext{10}{National Optical Astronomy Observatories, Tucson, AZ 85719, USA}
\altaffiltext{11}{Astronomy Department, University of Massachusetts, Amherst, MA 01003, USA}

\email{marco.castellano\char64inaf.it}
\begin{abstract}
We present deep spectroscopic follow-up observations of the Bremer Deep Field (BDF) where the two $z\sim$7 bright Ly$\alpha$ emitters (LAE) BDF521 and BDF3299 were previously discovered by \citet{Vanzella2011} and where a factor of $\sim$3-4 overdensity of faint LBGs has been found by \citet{Castellano2016}. We confirm a new bright Ly$\alpha$ emitter, BDF2195, at the same redshift of BDF521, $z=7.008$, and at only $\sim$90 kpc physical distance from it, confirming that the BDF area is likely an overdense, reionized region.  A quantitative assessment of the Ly$\alpha$ fraction shows that the number of detected bright emitters is much higher than the average found at z$\sim$7, suggesting a high Ly$\alpha$ transmission through the inter-galactic medium (IGM). However, the line visibility from fainter galaxies is at odds with this finding, since no Ly$\alpha$ emission is found in any of the observed candidates with $M_{UV}>$-20.25.
This discrepancy can be understood either if some mechanism prevents Ly$\alpha$ emission from fainter galaxies within the ionized bubbles from reaching the observer, or if faint galaxies are located outside the reionized area and bright LAEs are solely responsible for the creation of their own HII regions.
A thorough assessment of the nature of the BDF region and of its sources of re-ionizing radiation will be made possible by JWST spectroscopic capabilities.
\end{abstract}


\section{Introduction}
The redshift evolution of the fraction of LBGs showing Ly$\alpha$ emission \citep[e.g.,][]{Stark2010} allows us to put constraints on the Ly$\alpha$ transmission by the IGM. A substantial decrease of the Ly$\alpha$ fraction between z$\sim$6 and $z\sim$7 has been established by many independent analysis and interpreted as indication of a neutral hydrogen fraction $\chi_{HI}\sim$40-50\% at $z\sim$7 \citep[e.g.][]{Fontana2010, Vanzella2011, Pentericci2011,Schenker2012,Caruana2012,Pentericci2014}. The analysis of independent lines of sight presented in \citet{Pentericci2014}(P14 hereafter) has also shown that the decrease of the Ly$\alpha$ fraction suggests a patchy reionization process. 

Among the 8 pointings analysed by P14, the BDF \citep[][]{Lehnert2003} stands out as a peculiar area in the $z\sim$7 Universe. In fact, a single FORS2 slit mask observation of this field yielded the detection of two bright ($L\sim L^*$) Ly$\alpha$ emitting galaxies, namely BDF-3299 and BDF-521, at z=7.109 and z=7.008 respectively \citep[][V11 hereafter]{Vanzella2011}. These two objects, originally selected from our sample of VLT/Hawki-I z-dropout LBGs \citep[][C10b hereafter]{Castellano2010b}, show Ly$\alpha$ equivalent widths $>$ 50\AA~and are separated by a projected distance of only 1.9Mpc, while the  distance computed from Ly$\alpha$ redshifts is 4.4Mpc (see V11). The detection of bright Ly$\alpha$ emission from BDF-3299 and BDF-521 can be explained by these sources being embedded in an HII region that allows Ly$\alpha$ photons to redshift away from resonance before they reach the IGM \citep[e.g.][]{Miralda1998}. However, following \citet{Loeb2005} we estimated that these two galaxies alone cannot generate a large enough HII region, suggesting either the existence of additional ionizing sources in their vicinity \citep{Dayal2009,Dayal2011} or the contribution of AGN activity. 

We identified such potential, fainter re-ionizers through a follow-up HST program \citep[][C16a hereafter]{Castellano2016}. The dropout selection yielded a total of 6 additional highly reliable $z>$6.5 candidates at S/N(Y105)$>$10, corresponding to a number density $\gtrsim$3-4 times higher than expected on the basis of the z=7 UV luminosity function \citep{Bouwens2015,Finkelstein2015}. A stacking of the available HST and VLT images confirmed that these are robust $z\sim$7 sources.  A comparison between observations and cosmological simulations \citep{Hutter2014,Hutter2015} showed that this BDF overdensity has all expected properties of an early reionized region embedded in a half neutral IGM. 

In this paper we present deep spectroscopic follow-up of these additional LBGs aimed at estimating their Ly$\alpha$ fraction and redshift. If the BDF hosted a reionized \textit{bubble} we expect to measure a Ly$\alpha$ fraction higher than in average $z\sim$7 lines of sight and more consistent with the one measured at $z\sim$6. \textit{The BDF is the first $z\sim$7 field where a test of this kind can be performed}. The observations are described in Sect.~\ref{sect_obs}, while results and the estimate of the Ly$\alpha$ fraction are presented in Sect.~\ref{sect_res} and \ref{sect_lyafrac}, respectively. Finally, we discuss potential interpretations of our findings and directions for future investigations in Sect.~\ref{sect_disc}. Throughout the paper, observed and rest--frame magnitudes are in the AB system, and we adopt the $\Lambda$-CDM concordance model ($H_0$ = 70 km s$^{-1}$ Mpc$^{-1}$, $\Omega_M=0.3$, and $\Omega_{\Lambda}=0.7$).

\section{Observations}\label{sect_obs}
We observed the HST-selected candidates with FORS2 on the ESO Very Large Telescope and adopting the same setup used in our previous works which proved to be highly successful for confirming $z\sim 7$ galaxies. We used the 600z+23(OG590) grism (resolution R=1390), with slits $1''$ wide and a length in the range 6-12$''$. This setup maximizes the number of observed targets while enabling a robust sky subtraction and the maximum efficiency in the wavelength range $8000-10100$\AA.

Our primary targets in the BDF overdensity around the two Ly$\alpha$ emitters are the 6 S/N(Y105)$>$10 LBGs presented in C16a plus additional 8 sources at S/N(Y105)$\sim$5-10. All these sources have magnitude in the range Y105$\sim$26-27.3. We also re-observe the two bright emitters from V11. The mask was observed for a total of 29 hrs, resulting in 22.5 hrs of net integration time after excluding overheads and low quality frames.

\section{Results}\label{sect_res}
\subsection{A new confirmed emitter}
\begin{figure}[]
   \centering
 \plotone{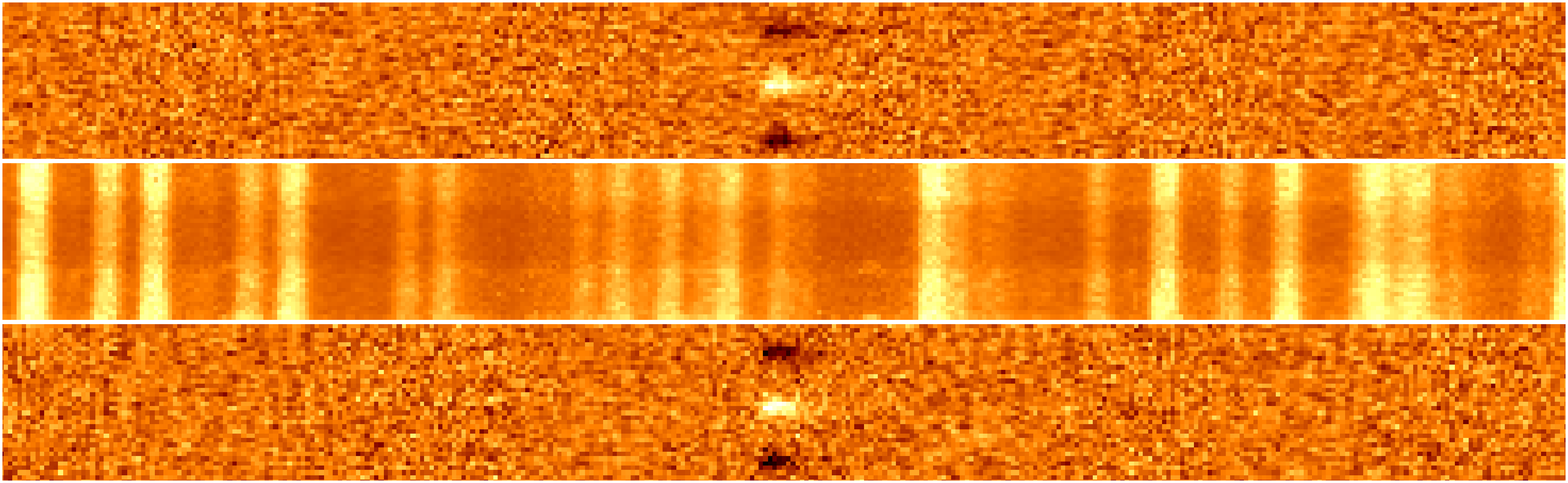}
 \plotone{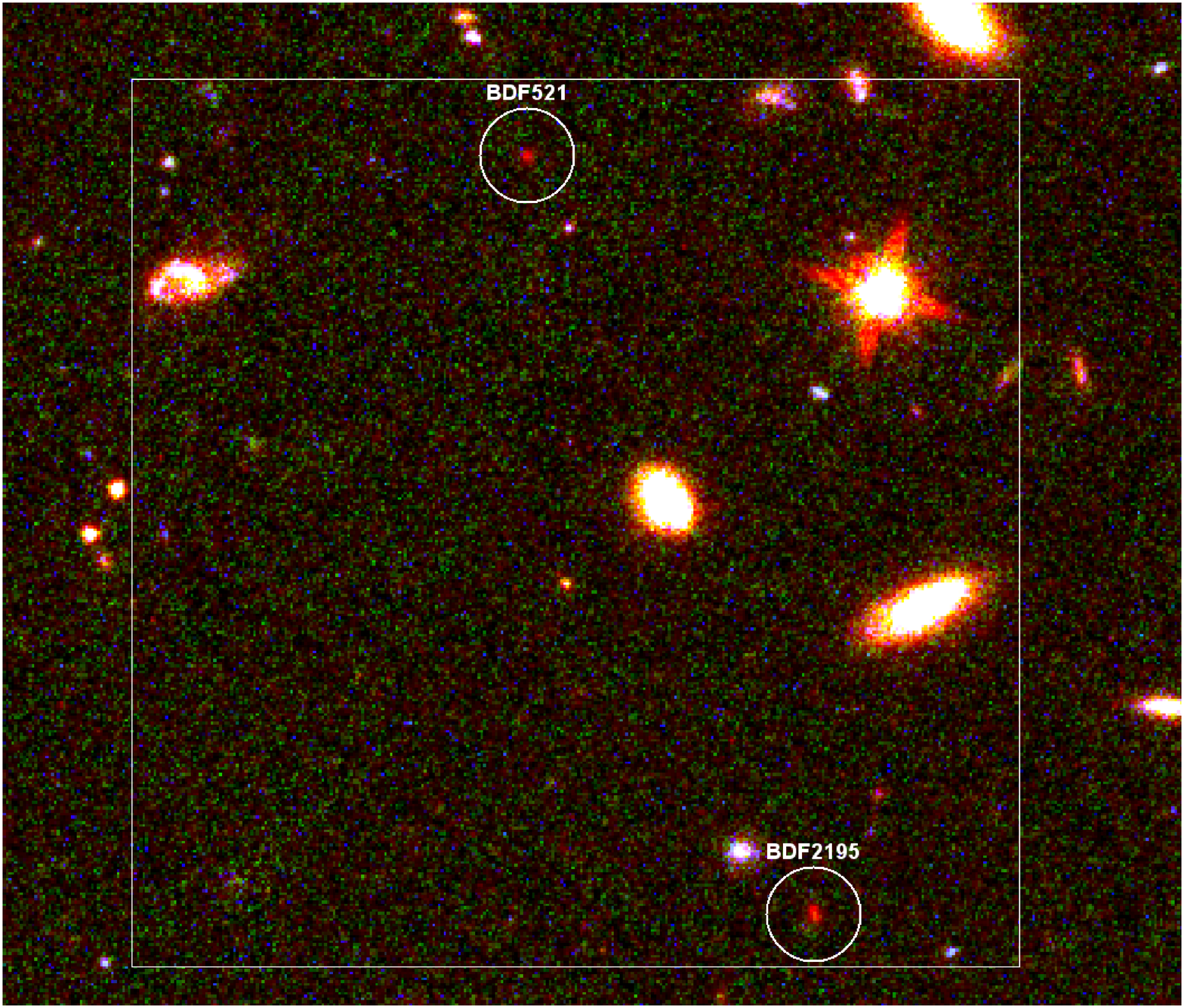}
   \caption{{\bf Top:} the FORS2 S/N spectra of BDF521 (top) and BDF2195 (bottom) together with the reference RMS spectrum (middle). {\bf Bottom:} the two objects on a HST three-color image of the BDF (blue, green and red channels are V606, I814 and Y105 respectively), the box side is of 100kpc (physical) at z=7.008.}\label{fig_spectra2d}%
\end{figure}

\begin{figure}[]
    \begin{center}    
\plotone{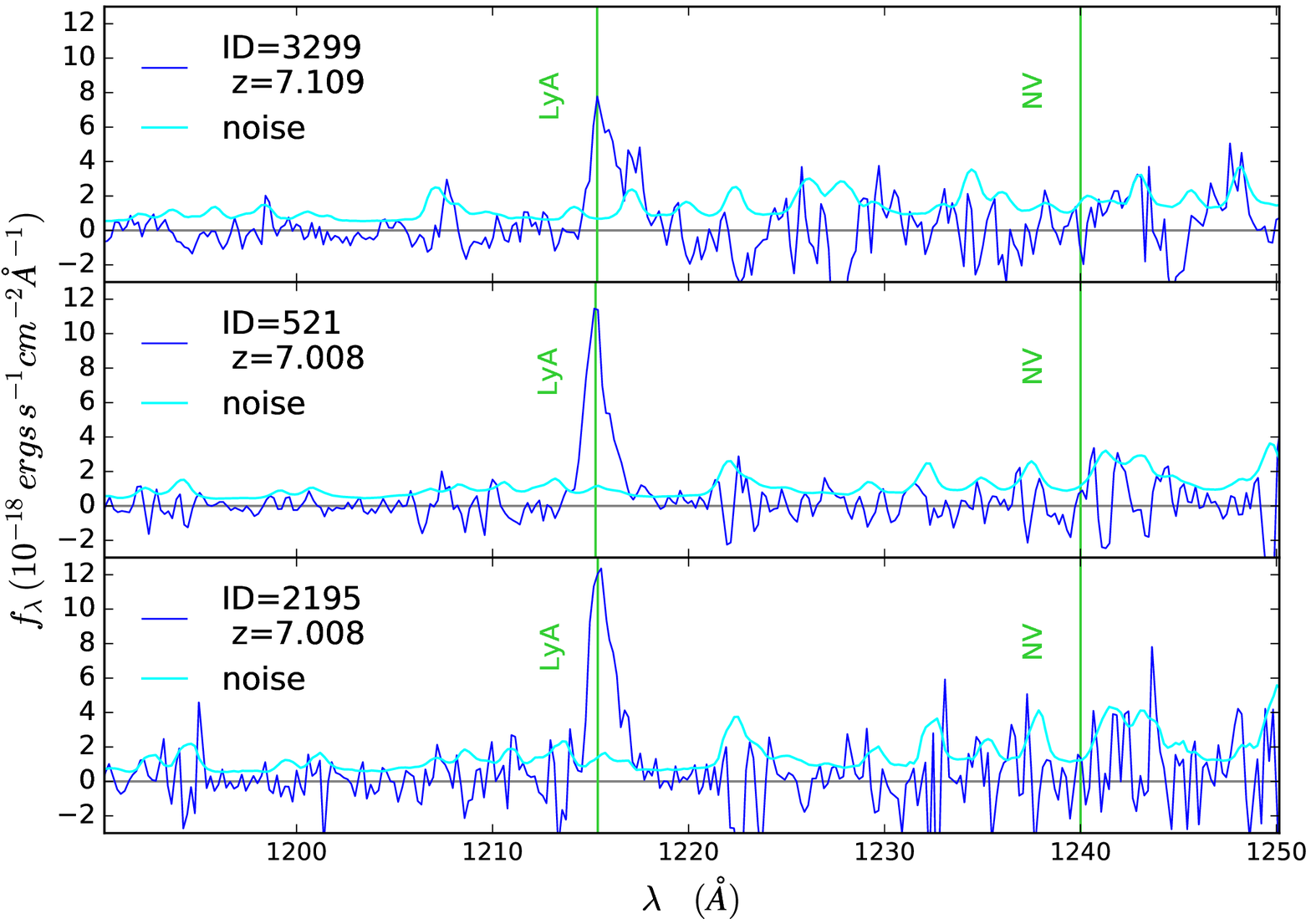}
\plotone{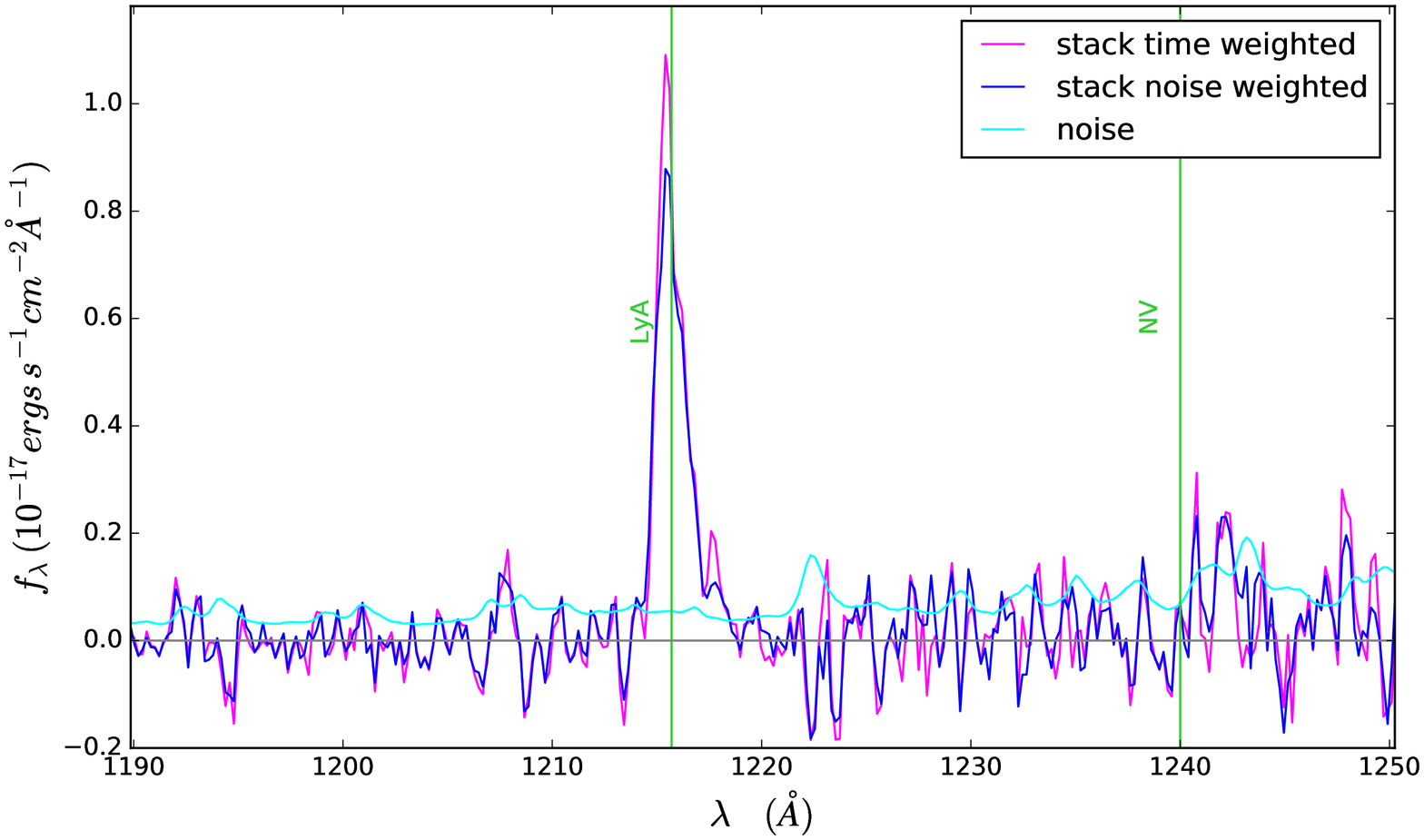}
\end{center}
\caption{\textbf{Top:} the three LAE in the BDF field: BDF521 and BDF3299 from Vanzella+11 (combining old and new FORS2 data), and the new one, BDF2195. \textbf{Bottom:} the stacked spectrum of the 3 emitters  obtained by scaling them with the total integration time applying a minmax clipping (magenta line), or weighting with the noise spectra (blue line, the resulting sky noise is shown in cyan).}\label{fig_spectra1d}%
\end{figure}

Out of the 16 candidates observed we confirm one new LBG with bright Ly$\alpha$ emission (Fig.~\ref{fig_spectra2d}), BDF2195 at mag Y105$\sim$26. This object was also detected in the HAWKI Y-band catalog presented in C10b but not included in the high-redshift sample due to photometric uncertainties.  We detect a clearly asymmetric Ly$\alpha$ line at $\lambda$=9737\AA~(z=7.008, Fig.~\ref{fig_spectra1d}), with FWHM$=$240 km/s (gaussian fit, corrected for instrumental broadening), and flux=1.85$\pm0.46\times$10$^{-17}$ erg s$^{-1}$ cm$^{-2}$, corresponding to an EW=50\AA. IGM absorption affects 22\% of this bandpass; accounting for this, and removing line contribution, the corrected Y105 continuum magnitude is 26.2. 

Intriguingly, BDF2195 has exactly the same Ly$\alpha$ redshift as BDF521, and the two have a projected physical separation of only 91.3kpc (Fig.~\ref{fig_spectra2d}). No additional lines are clearly found from spectra of the other observed candidates. To determine Ly$\alpha$ detection limits for these other objects, we adapted the simulations presented in F10, V11, P11, P14 and \citet{Vanzella2014} to the new observations (Sect.~\ref{subsect_sim}). The typical flux limit is $1.5 \times 10^{-18}$ erg s$^{-1}$cm$^{-2}$ in the range 8100 to 10000\AA~though it varies depending on the exact wavelength, due to the presence of bright sky emission in this spectral range. The corresponding rest-frame EW limit varies between 10 and 30\AA~across the redshift range z$\simeq$6-7.2. Details of the observed sample are reported in table~\ref{table1}. 

\begin{table*}
\begin{center}
\begin{tabular}{lccccccc}
\tableline
ID & RA & Dec &  $Y_{105}$  & Redshift &Ly$\alpha$ EW$^a$ (\AA) \\
\tableline
2883& 337.028076  &-35.160122 &25.97$\pm$0.08 &  - &$<$13 \\
2195 &336.942352 &-35.123257 &26.02$\pm$0.04  &  7.008$\pm$0.002  & 50$\pm$12 \\
401 &337.051239 &-35.172020 &26.43$\pm$0.08   & - & $<$11\\
3299 & 337.051147 &-35.166512 &26.52$\pm$0.08   & 7.109$\pm$0.002 & 50$\pm$6$^b$ \\
521 &336.944397& -35.118809&26.53$\pm$0.07   &7.008$\pm$0.002 & 64$\pm$6$^b$ \\
2009& 336.933716 & -35.124950& 26.89$\pm$0.14   & - & $<$17\\
994 &336.957092 &-35.136780 &27.11$\pm$0.19   & - & $<$19\\
1147&337.027130 &-35.163212& 27.26$\pm$0.11   & - &  $<$22\\
2660&336.940186& -35.116970& 27.27$\pm$0.10   & - & $<$22\\
2980 & 337.024994 &-35.142494&27.30$\pm$0.12   & - & $<$22\\
647 & 337.034332& -35.168716&27.31$\pm$0.15   & - & $<$23\\
1310& 336.953339 &-35.133030 & 27.32$\pm$0.16   & - & $<$23\\
2391&337.051361& -35.149185 & 27.33$\pm$0.17  & - & $<$23\\
187  & 336.953186 & -35.147457 & 27.33$\pm$0.10   &- & $<$23 \\
1899& 336.958618 & -35.126297 & 27.35$\pm$0.15  & - & $<$23\\
1807 &337.057861& -35.155842 &27.36$\pm$0.09  & - & $<$24\\
2192 & 337.018158 &-35.151600&27.40$\pm$0.10  & - & $<$25\\
\end{tabular}
\end{center}
\caption{FORS2 $z\sim$7 targets:  optical and spectroscopic properties}\label{table1}
\tablenotetext{a}{average 3$\sigma$ upper limits are computed at $7.008 \leq z \leq 7.109$}
\tablenotetext{b}{from V11}
\end{table*}

\subsection{Limits on NV$\lambda 1240$ emission}

The wavelength range observed by FORS2 covers the region of NV emission, where however no apparent emission signal is found in any of the three  Ly$\alpha$ emitters within 500 km/s from the expected position of the line \citep[e.g.][]{Mainali2018}, resulting in limits on the ratio Ly$\alpha$/NV$\gtrsim$8-10. We then built a weighted average spectrum of the three emitters (see Fig.~\ref{fig_spectra1d}) using all data of the present program and of our previous observations to compute limits on the NV emission, under the assumption that the shift between Ly$\alpha$ and NV emission is similar in the three sources. The stacked source has Ly$\alpha$ flux of $16.7\times 10^{-18}$ erg s$^{-1}$ cm$^{-2}$ and a NV$<3.36 \times 10^{-19}$ erg s$^{-1}$ cm$^{-2}$, corresponding to Ly$\alpha$/NV $>$ 17. This limit is much higher than the ratios measured in some z$\gtrsim$7 galaxies and considered indicative of AGN emission, ranging from Ly$\alpha$/NV$\sim$1-2 \citep{Tilvi2016,Sobral2017a,Hu2017} to $\simeq$6-9 \citep{Laporte2017,Mainali2018}. Our limit is also higher than the average Ly$\alpha$/NV$\sim$12 found in LBG-selected narrow-line AGNs at $z\sim$2-3 by \citet{Hainline2011}. However, the latter work also find that the Ly$\alpha$/NV distribution covers a wide range of values and Ly$\alpha$/NV $\gtrsim$20 are found \citep[see also][]{McCarthy1993,Humphrey2008}.  Finally, NV emission might also lack due to a very low metallicity \citep[though BDF3299 is already fairly enriched, as shown by ][]{Carniani2017}. It is thus not possible to rule out that our emitters also host AGN activity.

\section{The Ly$\alpha$ visibility in the BDF region}\label{sect_lyafrac}
\subsection{Simulations of the Ly$\alpha$ population at high-redshift}\label{subsect_sim}
Under the scenario where the BDF region is highly ionized compared to the average z=7 universe, our expectation was to detect Ly$\alpha$ also in several faint galaxies. Instead, we only confirmed one new bright source. To assess the significance of this result we run Monte carlo simulations to determine the expected number of objects we should have detected if the BDF region was similar in terms of Ly$\alpha$ visibility to the average z=7 Universe or to the average z=6 one (i.e., with a greater Ly$\alpha$ visibility). 
\begin{table*}[]
\begin{center}
\begin{tabular}{lccc}
\tableline
Sample &Total & Bright & Faint       \\
\tableline
Observed        &             17     & 5        & 12 \\
\tableline
Detected in Ly$\alpha$     &               3     & 3        & 0 \\
\tableline
\end{tabular}
\end{center}
\begin{center}
\begin{tabular}{lcccccccc}
\tableline
PDF(z) & Ly$\alpha$   &  & Probability &  & & Expected Number &   \\
  &  visibility &$P(tot=3)$&$P(bright=3)$&$P(faint=0)$&$<N_{tot}>$&$<N_{bright}>$&$<N_{faint}>$\\
\tableline
\tableline
 Flat  &  z=7 & 0.21 & 0.009 & 0.17& 2.1  &0.7  &1.4  \\
 P(z,Y)           &  z=7 &  0.18& 0.009& 0.22&1.9 &  0.7  & 1.2  \\
 Flat   &  z=6 & 0.08& 0.035& 0.002 &5.5  & 1.2  &4.3  \\
 P(z,Y)          &   z=6 &  0.11& 0.036&  0.004&5.0     & 1.2  & 3.8  \\

\tableline
\end{tabular}\caption{\textbf{Top}: the observed number of bright and faint candidates, assuming sources to be at z=7 ($M_{UV}<-20.25$ corresponding to $Y_{105}<$26.7). \textbf{Bottom:} probability of observed 3$\sigma$ Ly$\alpha$ detections, and expected number of detected lines for the total, bright and faint subsamples, as computed from Monte Carlo simulations. }\label{table2}
\end{center}
\end{table*}
We consider the 17 sources presented in Table~\ref{table1}, namely the 16 targets discussed in Sect.~\ref{sect_obs} plus another bright object (BDF2883 at Y=25.97) that was observed in the same region with the old FORS2 mask (P11 and V11). First, for each object without a confirmed redshift we extract randomly a redshift according to two cases: a) we assume that the redshift distributions of the candidates follow those derived from the LBG color selection from C16a. These distributions ($P(z,Y)$ in Table~\ref{table2}) are derived from simulations for three $\Delta mag=0.5$ bins at Y=26-27.5, and peak at $z\sim$6.9 with magnitude-dependent tails covering the range $z\sim$6.0-7.8;  b) we assume a \textit{flat} redshift distribution in the small redshift range [6.95:7.15] approximately corresponding to a size of 10 Mpc, thus assuming all sources to be part of a unique, localized structure. 

When a redshift is assigned, we calculate the rest-frame $M_{UV}$ (at $\lambda=$1500\AA) of the galaxy on the basis of the observed magnitude assuming a flat spectrum, and we determine the limiting flux at 3$\sigma$ from the expected position of the Ly$\alpha$ line. We then calculate the limiting Ly$\alpha$ line EW ($EW_{lim})$ on the basis of the limiting flux and the observed magnitude. For all objects with a Ly$\alpha$ detection we fix the redshift at the spectroscopic value and determine $M_{UV}$ and $EW_{lim}$ as above.  The observed continuum flux is computed from the observed Y105 magnitude for all sources with no line detection. In the case of BDF521 and BDF2195 we adopt as reference the J125 magnitude \citep{Cai2015} which samples the UV at 1500\AA~and is not affected by IGM and Ly$\alpha$ emission, while for BDF3299 we correct the observed Y105 magnitude by subtracting line emission and accounting for the portion of filter (27\%) sampling IGM.
We extract an intrinsic Ly$\alpha$ EW ($EW_{intr}$) for our object: this is randomly drawn from the observed EW distributions which are derived separately for faint and bright galaxies ($M_{UV}<-20.25$) from more than 160 $z\sim$6-7 LBGs in the CANDELS fields \citep[][and Pentericci et al. submitted]{DeBarros2017,Castellano2017}.
If $EW_{intr}> EW_{lim}$ then the galaxy is counted as a detection, otherwise it is counted as a nondetection.  If the extracted redshift is beyond the FORS2 range ($z\lesssim$7.3), it is automatically counted as a nondetection. The procedure is repeated 10$^5$ times for all 17 input objects. As additional input parameter, we can allow a fraction of the objects to be undetected because they are lower redshift interlopers.

We obtain: 1) the fraction and total number of expected detections at bright and faint magnitudes; 2) the probability of having a total of 0,1,2 etc detections in each sample, and 3) the EW distribution of the detected objects.

\subsection{Prevalence of bright Ly$\alpha$ emitters}

We find that the number of bright detected objects (3) points to a very high line visibility in this region. In fact, it is higher than expected at $z\sim$7 and even higher than the $z\sim$6 statistics, though more consistent with the latter scenario of a ``clean'' $z\sim$6-like visibility: the probability of finding 3 bright emitters given the known $z\sim$7 Ly$\alpha$ visibility is less than 1\%. However, the Ly$\alpha$ fraction among faint galaxies is strikingly at odds with the expectations for the ``reionized'' case. We expect $\sim$4 detections in the faint sample for a $z\sim$6 EW distribution, and the probability of finding none is 0.2-0.4\%. No appreciable difference is found among the ``flat'' and ``P(z,Y)'' cases.  The results are summarized in table~\ref{table2}.

Adopting a 5$\sigma$ threshold, the number of expected detections decreases by 25\%-35\% for bright sources and 40-100\% for faint ones, depending on the redshift distributions, but remains inconsistent with the observations.

We have so far assumed that all our sources are genuine high-redshift galaxies. Only under the extreme assumption of a $\gtrsim$50-70\% fraction of interlopers at faint end (at 5$\sigma$ and 3$\sigma$ thresholds for line identification respectively), the null Ly$\alpha$ detection rate among faint galaxies can be reconciled with a $z\sim$6 EW distribution at both bright and faint fluxes. We consider this a very unlikely possibility given the conservative selection criteria adopted and the fact that in none of the unconfirmed galaxies do we detect other features that could point to a low redshift nature.  

\section{Discussion and conclusions}\label{sect_disc}
The high detection rate of Ly$\alpha$ emission in the BDF bright sources supports the scenario from C16a, namely the BDF hosts a reionized bubble where Ly$\alpha$ visibility is enhanced. However, the lack of Ly$\alpha$ detections in faint galaxies is apparently at odds with such a picture. 

Our observations could imply that contrary to the reference scenario outlined in C16a, the faint galaxies are actually outside the bubbles, while the bubbles are created by the bright galaxies alone, or thanks to the contribution of objects beyond the current detection limit \citep[as the z$\sim$6 clustered ultra-faint dwarfs observed by][]{Vanzella2017a,Vanzella2017b}. The faint galaxies might be part of a superstructure which includes the reionized regions, but their Ly$\alpha$ might be undetected because they lie outside the patches with low neutral fraction. Unfortunately the available HST imaging observations do not cover the full BDF region ($\sim 2.4\times 2.4$ Mpc at $z \sim$7) but only two $\sim 0.7 \times 0.7$ Mpc areas centred on the emitters, thus preventing detailed constraints on the extent and geometry of the overdensity.

\begin{figure}[!ht]
   \centering
   \epsscale{1.2}
\plotone{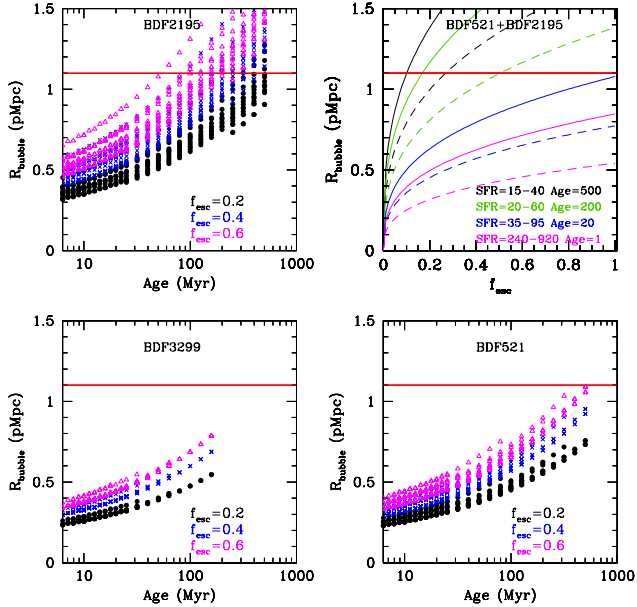}
\caption{The size ($R_{bubble}$) of the ionized bubbles created by the three emitters, as a function of the age of the stellar population (bottom panels and top-left panel) for BPASSV2 SED models within 68\% c.l. from the best-fit. The cases for escape fraction $f_{esc}$=0.2,0.4, 0.6 are shown as black circles, blue crosses and magenta triangles respectively. Solutions having the same age and $f_{esc}$ differ in terms of either best-fit metallicity or E(B-V), or both. The top-right panel show minimum (dashed lines) and maximum (continuous lines) $R_{bubble}$ as a function of $f_{esc}$ adding contribution from BDF521 and BDF2195, for stellar populations of age 500 (black), 200 (green), 20 (blue) and 1 Myr (magenta). The horizontal red line in all plots mark the minimum HII size $R_{min}$=1.1Mpc enabling Ly$\alpha$ to escape.}\label{fig_bubbles}%
\end{figure}

To ascertain whether the BDF emitters are capable of re-ionizing their surroundings or not, we performed SED-fitting on the available photometry (see C16a for details) and estimated the SFR and ionizing flux of the BDF emitters with our $\chi^{2}$ minimization code \textit{zphot.exe} \citep{Fontana2000}. The SFR and the age of the galaxies are then used to measure the size ($R_{bubble}$) of the resulting ionized bubbles  assuming a hydrogen clumping factor C=2 and an average neutral hydrogen fraction $\chi_{HI}=0.5$ surrounding the sources at the onset of star-formation \citep[see, e.g.,][]{Shapiro-giroux1987,Madau1999}. We used both BC03 \citep{Bruzual2003} and BPASSV2.0 \citep{Eldridge2009,Stanway2016} templates with constant SFR, age from 1Myr to the age of the universe at the given redshift, E(B-V) in the range 0.0-1.0 \citep[assuming][extinction curve]{Calzetti2000} and metallicity from 0.02$Z_{\odot}$ to solar. In Fig.~\ref{fig_bubbles} we show the $R_{bubble}$ of the ionized regions created by BPASS SED models within 68\% c.l. from the best-fit for the three emitters, as a function of the age of the stellar population and for different values of the $f_{esc}$. We also show the $R_{bubble}$ ranges for the case where we summed together the ionizing fluxes of the two sources BDF521 and BDF2195 that form a close pair at only $\sim$90 kpc projected speration. The size $R_{bubble}$ must be compared to the dimension $R_{min}$=1.1Mpc \citep[estimated as in][]{Loeb2005} enabling Ly$\alpha$ to be redshifted enough to reach the observers. 

On the one hand, BDF521 and BDF2195 would require a high $f_{esc}\gtrsim$20-60\% to create a large enough bubble, while BDF3299 is unable to create its own bubble even assuming 100\% escape fraction. On the other hand, when summing the two contributions the BDF521-BDF2195 pair can create a large enough bubble with $f_{esc}\gtrsim$10-15\% (BC03 and BPASSV2 respectively) and constant star-formation for $\gtrsim$400Myr. We do not find solutions that allow age $<$20Myr, which is also consistent with supernovae requiring 3.5-28 Myrs to build channels that can allow LyC photons to escape \citep{Ferrara2013}.  We find that results obtained with BPASSV2 library point to slightly higher re-ionizing capabilites compared to BC03 ones, as slightly smaller fitted SFRs partially compensate the higher ionizing photons production rate of the BPASS models. The aforementioned $R_{min}$ assumes that the Ly$\alpha$ escapes from the galaxies at the systemic redshift. However, line visibility from smaller HII regions is possible in presence of strong outflows: a 220km/s shift, which is the median value for galaxies in massive halos from \citet{Mason2018}, results in $R_{min}\sim 0.85$ Mpc, which can be reached in a few 100Myr by the BDF521-BDF2195 pair with $f_{esc}\lesssim$10\%, but still out of reach for BDF3299 without extreme $f_{esc}$.

The case described above considers star formation as the only source of ionizing photons. However, we cannot exclude that the BDF emitters host AGN that could provide a substantial contribution to the ionizing budget, or that the bubbles have been created by past AGN activity. In such a case, bright emitters including BDF3299 could be solely responsible for the creation of reionized regions, assuming lower $f_{esc}$ and/or ages for their stellar populations.

As an alternative to scenarios where the ionizing flux is generated by bright galaxies alone, some mechanism must be in place to prevent Ly$\alpha$ from faint galaxies to reach the observers. A possibile explanation can be found in an accelerated evolution of overdensity members compared to the normal field population. The bright emitters are young, relatively dust-free sources \citep[consistent with the ALMA results from][]{Maiolino2015} experiencing a bursty episode of star-formation. Intense bursts of star-formation favoring the escape of Ly$\alpha$ photons are stimulated by an enhanced rate of mergers and interactions within the overdensity. In this picture, all faint LBGs are actually more evolved objects, thus with intrinsically fainter line emission, that have already experienced such bursty star-formation episodes in the past. Recombination of neutral hydrogen in the regions close to overdensity members can provide an additional mechanism explaining lack of line emission from faint galaxies, as only in bright galaxies with large circular velocities Ly$\alpha$ photons acquire a frequency shift enabling their escape from the circum-galactic medium.

Indeed, as discussed by \citet{Mason2018}, Ly$\alpha$ emission from UV bright galaxies residing in reionized overdensities can be further boosted by their higher velocity offsets that reduce the damping wing absorption by cosmic neutral hydrogen. This effect, possibly along with enhanced  Ly$\alpha$ photon production, has been proposed as a physical explanation for the increased Ly$\alpha$ visibility in very bright ($M_{UV}<$-22) $z>$7 galaxies found by \citet{Stark2017}. While the three BDF emitters at $M_{UV}\gtrsim$-21 are not as bright, the combination of a large enough HII region around them and of frequency shifts induced by their circular velocities likely plays a role at enhancing their Ly$\alpha$ visibility with respect to $z\sim$6 LBGs.

Luckily, a thorough examination of the aforementioned scenarios will soon be made possible by observations with JWST. It will be possible to: 1) confirm a very low neutral fraction in the region surrounding the bright emitters by looking for blue wings in high-resolution Ly$\alpha$ spectra \citep[e.g.][]{Hu2016}; 2) clarify the nature of bright emitters through a more accurate measurement of SFR, extinction and age (H$\alpha$ luminosity, H$\alpha$/H$\beta$ and H$\alpha$/UV ratios), and probing signatures of a high escape fraction \citep[EW of Balmer lines or the $O_{32}$ ratio, e.g.,][]{Castellano2017,DeBarros2016,Chisholm2018}, AGN emission and hard ionizing stellar spectra \citep[e.g.,][]{Mainali2017,Senchyna2017}; 3) assess whether faint candidates are members of a localized overdensity at $z\simeq$7.0-7.1 as the bright ones, or just outside such a region, or low-z interlopers in the sample by measuring their redshift from optical emission lines; and 4) measure velocity shifts between Ly$\alpha$ and UV/optical lines that trace the systemic redshift of bright emitters. 

A systematic analysis of this kind carried out with JWST on $z\gtrsim$7 lines-of-sight with different levels of Ly$\alpha$ visibility will eventually shed light on the processes responsible for the creation of the first reionized regions.

\acknowledgments
Based on observations collected at the European Organisation for Astronomical Research in the Southern Hemisphere under ESO programme  099.A-0671(A).
PD acknowledges support from the European Research Council's starting grant ERC StG-717001 and from the European Commission's
and University of Groningen's CO-FUND Rosalind Franklin program.

\bibliographystyle{aasjournal}

\end{document}